\documentstyle[11pt,IAUS215,twoside]{article}

\def\edcomment#1{\iffalse\marginpar{\raggedright\sl#1\/}\else\relax\fi}
\reversemarginpar

\begin{document}
\newcommand{\demi}{\frac{1}{2}}
\newcommand{\vv}{\vec{v}}
\renewcommand{\vr}{\vec{r}}
\newcommand{\vF}{\vec{F}}
\newcommand{\dt}[1]{\frac{\partial  #1}{\partial t}}
\newcommand{\dphi}[1]{\frac{\partial  #1}{\partial \varphi}}
\renewcommand{\na}{ \vec{\nabla} }
\newcommand{\lp}{ \left(}
\newcommand{\rp}{ \right)}
\newcommand{\eps}{\varepsilon}
\newcommand{\dnt}[1]{\frac{d  #1}{dt}}
\newcommand{\vO}{\vec{\Omega}}

\vspace*{1cm}
\title{Evolution of rotation in binaries: physical processes}
\author{Michel Rieutord}
\affil{Observatoire Midi-Pyr\'en\'ees, 14 avenue Edouard Belin, 31400
Toulouse, France}

\begin{abstract}
In this review, we describe the physical processes driving the dynamical
evolution of binary stars, namely the circularization of the orbit and
the synchronization of their spin and orbital rotation. We also discuss
the possible role of the elliptic instability which turns out to be an
unavoidable ingredient of the evolution of binary stars.
\end{abstract}

\section{Introduction}

The evolution of rotation is usually associated with an evolution of
angular momentum; changing the angular momentum of any body requires
torques and stars do not escape from this law of physics. In binary
stars there is a permanent source of torques: the tides. Hence,
understanding the evolution of rotation of stars in a binary system
demands the understanding of the action of tides. This compulsory
exercise was started more than thirty years ago by Jean-Paul Zahn during
his ``Th\`ese d'\'etat", {\it Les mar\'ees dans une \'etoiles double
serr\'ee} (Zahn 1966). All the concepts needed to understand tides and
their actions in the dynamical evolution of binaries are presented in
this work.

\subsection{Why should we consider the rotation in binary stars ?}

Surely, as in isolated stars, rotation is an important ingredient of
evolution through the induced mixing processes: turbulence in stably
stratified radiative zones, circulations... All these processes will
influence the abundances of elements in the atmospheres or the internal
profile of angular velocity, for instance.

However, in binary stars new phenomena appear: tides. They make the orbit
evolving, force some mixing processes (through eigenmode resonances for
instance) or may even generate instabilities leading, to some
turbulence (see below).

These new phenomena need also to be understood if one wishes to decipher
the observations of binary stars. In this respect binary stars offer
more observables than single stars like the parametres of the orbit, masses of the stars, their
radii, etc. If the system has not exchanged mass during
evolution and if other parameters like luminosity, surface gravity,
abundances can also be determined unambiguously, binary stars offer new
constrains on the stars which may be useful for our understanding of
stellar evolution. Also, a statistical view of orbital
parameters may constrain the dissipative processes at work in these
stars (Mathieu et al. 1992).

\subsection{Synchronisation and circularization : how it works.}

Let us consider an isolated system made of two stars of mass M$_1$,
M$_2$, of moment of inertia I$_1$, I$_2$ and of spin angular velocity
$\vec\Omega_1$, $\vec\Omega_2$. The semi-major axis of the orbit is $a$
and the eccentricity $e$. For simplicity we shall assume that the
angular momentum vectors are all aligned. Hence, the total (projected)
angular momentum of
the system, which is conserved during evolution, reads:

\begin{equation} L= \frac{M_1M_2}{M_1+M_2}a^2\Omega_{\rm orb}\sqrt{1-e^2} +
I_1\Omega_1 + I_2\Omega_2 
\end{equation}
On the other hand, the total energy of the system, namely,

\[ E=-\frac{GM_1M_2}{2a} + \demi I_1\Omega_1^2 + \demi I_2\Omega_2^2\]
decreases because of dissipative processes.

To appreciate the natural evolution of such a system, let us consider
the even more simplified system where the angular momentum and the
energy of the spin of the stars are negligible compared to their orbital
equivalent. Using Kepler third law to eliminate the mean angular velocity of
the orbital motion $\Omega_{\rm orb}$, the previous equations lead to

\[ L_{\rm system} \sim a^{1/2}\sqrt{1-e^2}\quad {\rm and}\quad E_{\rm
system} \sim -\frac{GM_1M_2}{2a} \]
During evolution the system loses energy through dissipative mechanisms,
thus $a$ decreases which implies that $e$ also decreases to insure a
constant angular momentum. Thus, with time the orbit slowly circularizes.

Once the orbit is circular or nearly circular, the system may continue
to evolve if the orbital angular velocity and spin angular velocity are
not identical: this is the synchronization process after which the system
has reached its minimum mechanical energy state: all the stars rotate
at the same rate, i.e.  $\Omega_{\rm spin} = \Omega_{\rm orb}$
like the moon on its terrestrial orbit.

\subsection{Tides}

In the foregoing section we presented a global view of the evolution of
the system, however the way the orbit or the spin change is controlled by
the torques applied to the stars. As we said at the beginning, a
permanent source of torques is given by the tides which therefore need
to be studied. But what is a tide?

The tide is the fluid flow generated by the tidal potential, i.e. the
azimuth dependent part of the gravitational potential inside a star. In
short, if you sit on a star, it is the forced flow generated by the
celestial bodies orbiting around you. If you sit on Earth you feel the
tides of the moon and the sun essentially.

Now let us assume that the tidal forcing is milde enough so that the
fluid flow obeys linear equations; formally, we may write the system
like

\[ \rho\dt{\vv} + {\cal L}(\vv) = \vF_T({\vr})f(t) \]
where we assume that the tidal force $\vF_T({\vr})f(t)$ can be separated
into its spatial and temporal dependence. Written in this way we
immediately see that if the inertia of the fluid can be neglected (i.e.
the term $\rho\dt{\vv}$), then the velocity field can be computed with
the same temporal dependence as the exciting force. The response is
instantaneous. Moreover, if Coriolis acceleration and viscosity are
negligible, the only response of the fluid is through a pressure
perturbation, i.e. it is purely hydrostatic. This extremely simple, but
not unrealistic, case is called the {\em equilibrium tide}.

On Earth, this tide is enough to understand the basic properties of
terrestrial tides: i.e. that there are two tides a day, that their
amplitude is $\sim50$~cm or that they are stronger at full or new moon;
the hydrostatic perturbation describes the famous tidal bulge which is
often representing tides in elementary courses. Such a description is
appropriate if you sit on the mediterranean beaches or on those of the
Pacific ocean; however, if you sit on the Atlantic shore, like here in
Cancun, you will easily notice that the tide is much larger than the
expected 50~cm. In the Mont Saint-Michel bay it easily reaches 10 meters !

The difference comes from what we neglected: the inertia of the fluid
and the ensuing resonance phenomenon. For the tidal wave, whose
wavelength is a few thousand kilometers, the Atlantic ocean is a mere
puddle five kilometers deep. Surface gravity waves may thus be studied
using the shallow water approximation and their speed is given by 

\[ V_{\rm wave} = \sqrt{gh} \simeq 220\;{\rm m/s}\]
where $g$ is the gravity and $h$ the depth. With a mean width of
5000~km, the Atlantic is crossed twice in 12.6 hours; but the tidal
forcing is back after 12.4 hours. Obviously, we are close to a resonance
and in this case the equilibrium tide is insufficient to describe
the tidal response of the fluid. In this case the tide will be qualified
as {\em dynamical}.

Quite clearly, the equilibrium tide is much easier to handle than the
dynamical one; this is why it was first studied (Zahn 1966); the
dynamical tide received first serious considerations by Zahn (1970) and
a proper treatment by Zahn (1975).

\subsection{Torques and tidal evolution}

In order to compute the dynamical evolution of the system, namely the
parameters of the orbit and the spin of the stars it is necessary to
evaluate both the torques and the dissipation inside the stars.

The torque suffered by a star results from an unsymmetrical distribution
of mass with respect to the tidal potential; mathematically, 

\[ {\cal T}_z = -\int_{(V)} \lp\vr\times\rho\na\Phi_T\rp_zdV =
-\int_{(V)}\dphi{\Phi_T}\rho' dV \]
where $\rho'$ is the density perturbation generated by the tidal
potential $\Phi_T$; $\varphi$ is the angular azimuthal variable.
We see from this expression that
torques can only exist if the excitation $\Phi_T$ and the response
$\rho'$ are out of phase (or antiphase). The phase lag between these two
quantities comes from the dissipative mechanisms at work in the stars
and is usually a small but important quantity.

From the expressions of the energy and angular momentum of the orbit, we
can derive:

\[ \dnt{\ln a} = \frac{\dot{E}_{\rm orb}}{|E_{\rm orb}|} \]
\[ \dnt{e^2} = {|E_{\rm orb}|}^{-1}\lp (1-e^2){\dot{E}_{\rm orb}} -
\Omega\sqrt{1-e^2}\dot{L}_{\rm orb}\rp \] 
where $\dot{E}_{\rm orb}$ and $\dot{L}_{\rm orb}$ are respectively the
power dissipated in the stars and the torques exerted on the orbit (the
parallelism of angular momentum vectors is still assumed).

The foregoing discussion therefore shows that, provided the stars do not
change, all the evolution of the system is controlled by the dissipation
of energy by the tidal flow.

In the case stellar evolution is important, for instance before or after
the main sequence, changes in the inertia momenta will change the spin
of the stars and the applied torques.

\section{Dissipation mechanisms}

\subsection{Viscosity}

The first and most obvious physical mechanism to dissipate mechanical
energy is viscosity. However, the viscosity of stellar plasma is far too
weak to be efficient and only turbulent viscosity of convection zones
can significantly affect the evolution  of the system.

But the effectiveness of turbulent viscosity is hampered by the
fact that tidal flows are periodic in time: in any region of a
convective zone where the lifetime of eddies is longer than the period
of the tidal flows, the turbulent eddy viscosity is reduced. The
question is, of course, how much it is reduced compared to its usual
approximation

\[ \nu_T^0 = \frac{1}{3}V_{\rm turb}\ell_{\rm turb}\]
where $V_{\rm turb}$ and $\ell_{\rm turb}$ are respectively the
velocity of turbulent eddies and their mean free path. 

Presently, two prescriptions coexist: the first by Zahn (1966) says that

\[ \nu_T = \nu_T^0 {\rm min}\lp 1,\frac{V_{\rm turb} P_{\rm
tide}}{2\ell_{\rm turb}} \rp
\]
which means that when the period of tides is shorter than the turnover
time of the eddies, the mean-free path of the eddies should be reduced
to the distance covered by an eddy during half a period.

The second prescription is originally due to Goldreich and Keeley (1977)
but adapted by Golman and Mazeh (1991); it  says that

\[ \nu_T = \nu_T^0 {\rm min}\lp 1,\lp\frac{V_{\rm turb} P_{\rm
tide}}{2\pi\ell_{\rm turb}} \rp^2\rp
\]
This prescription means that the turbulent viscosity should be such that

\[ \nu_T \lp\frac{V_{\rm tide}}{\ell_{\rm tide}}\rp^2=<\epsilon> \]
namely that the dissipation by tidal currents fits the dissipation by
the turbulent cascade $<\eps>$ (recall that in the Kolmogorov cascade
$<\epsilon>\sim V_\ell^3/\ell$).

These two prescriptions yield different exponent in the dependence of the
evolution time scale with the period of the system; in principle, they can
be discriminated by observations. However, this exercise turns out to be
difficult; moreover, using binaries in stellar clusters, Mathieu et al.
(1992) found that no reduction of viscosity was necessary to explain
observations ! Obviously, more work is needed to clarify this question.

Another viscous damping process was also put forward by Tassoul (1987),
namely the dissipation in Ekman boundary layers. However, we have shown
that because of the stress-free surfaces of the stars, such layers are
rather regions with lower dissipation than the rest of the star (see
Rieutord 1992, Rieutord and Zahn 1997).

\subsection{Radiative damping}

Another way to dissipate energy of tidal flows is through radiative
damping. This mechanism will affect essentially radiative zones; indeed,
from the tidal excitation, one needs to generate temperature
fluctuations which are dissipated by radiative diffusion.

The most natural way to generate these temperature fluctuations comes
from the excitation of gravity modes since, mechanically, they are
associated with the buoyancy force. It is
therefore quite clear that the dissipation and the ensuing torques will
be most important when the forcing frequency is near that of an
eigenmode of the star. From this remark,
it is also clear that only the dynamical tide will be relevant in this
process.

On general grounds we may observe that the tidal forcing is
low-frequency and that low-frequency gravity modes are high order modes
which can then be described by an asymptotic theory. This was the way
chosen by Zahn (1975) and later by Goldreich and Nicholson (1989).

This mechanism is essentially relevant for early-type stars which own an
outer radiative zone. As shown by the previous authors the  tidal
excitation is most intense near the core-envelope boundary but since
gravity waves are only partially reflected at the star surface, they
deposit their angular momentum there and these layers are synchronized
first. This argument was developed by Golreich and Nicholson (1989) to
explain the higher synchonization rate observed in early-type stars compared
to the theoretical predictions of Zahn (1977).

These studies are based on an asymptotic approach and ignore the role of
the Coriolis force. This force is, however, unavoidable since the stars are
always rotating and in most cases the tidal forcing is also in the band
of the so-called 'inertial modes' whose restoring force is precisely the
Coriolis force. In fact inertial modes always combine with gravity modes
as they are also low-frequency modes; both together, they occupy the band $[0,{\rm
max}(2\Omega_s,N_{\rm max})]$, where $N$ is the Brunt-V\"ais\"al\"a
frequency.

Recently, much progress was achieved in the direction of including the
whole spectrum of low-frequency modes in the tidal response of an
early-type star. Indeed, Witte and Savonije (1999a,b,2001) computed
numerically the response of a 10 M$_\odot$ main sequence star and the ensuing
evolution of the system with various inital eccentricities. Their
results show many interesting features:
\begin{itemize}
\item The efficiency of resonance crossing at decreasing the
eccentricity when it is initially small (a few percent).
\item A new phenomenon, which they called ``resonance locking" and by
which two resonant modes, one tending to spin-up the star and the other
trying to spin it down, yield equilibrating torques which maintain the
two stellar modes close to resonance and therefore force a strong evolution
of eccentricity (see figure 3 in Witte and Savonije 1999b).
\end{itemize}

The calculations of Witte and Savonije show that radiative damping is
efficient at reducing the eccentricity of the orbit in  a fraction of the
lifetime of the star: in their examples, the 10 M$_\odot$ star has a
lifetime of 20~Myrs; an initial eccentricity of 0.02 is erased in 1~Myr,
but an initial eccentricity of 0.25 or 0.7 reduces to 0.1 in 3~Myrs; a
further evolution apparently operates on a much longer time scale.

Obviously, if the model of  Witte and Savonije could be used as a true
system, it would mean that orbits with $e=0.1$ may be as evolved as $e=0$
ones and differ only by initial conditions.

\subsection{A new possibility: the elliptic instability}

Likely, all the mechanisms by which energy can be dissipated have not
been examined. We shall now present a new one caused by the elliptic
instability and which is potentially a rather strong source of
dissipation.

The elliptic instability has been studied mainly in simple geometries
and even in these cases it is a difficult problem. For a recent review we
refer the reader to the work of Kerswell (2002). Presently, the work closest to
astrophysical applications is that of Seyed-Mahmoud et al. (2000) who
investigated this instability in an ellipsoidal configuration for the
core of the Earth.

The basic result we need from fluid mechanics is the growth rate $\gamma$
of this instability, namely

\[ \gamma \sim \varepsilon\Omega\]
where $\varepsilon$ is the ellipticity of streamlines and $\Omega$ the
rotation rate of the vortex.

Schematically, this instability may be seen as a parametric
instability: the solid body rotating fluid feels a perturbation (the
ellipticity of the stream lines) with a frequency $2\Omega$. Such a
periodic forcing can destabilize modes at half its frequency namely
$\Omega$; this is precisely the frequency of the so-called spin-over
modes (or Poincar\'e modes). Such modes are solid-body rotation  around
an axis in the equatorial plane. When the instability develops and the
amplitude gets sufficiently large, the vortex start to precess and is
usually completely destroyed as observed in experiments (Malkus 1989).
Hence, this instability may be very important as it is able to dissipate the
total kinetic energy of the vortex. But let us consider the
astrophysical case.

For the sake of simplicity we consider a main sequence star in a binary
system with a circular orbit. The tidal perturbation is provided by a
point-mass object. The system is not synchronized and therefore
$\Omega_{\rm spin}\neq\Omega_{\rm orb}$. In a reference frame rotating
with the tidal potential, the main sequence star is like a strained
vortex rotating with the angular velocity $\Omega_{\rm spin}-\Omega_{\rm
orb}$ and enduring an elongation $\eps$ of the tidal potential.

In such a case the instability develop on a time scale
$\lp\,\eps(\Omega_{\rm spin}-\Omega_{\rm orb})\,\rp^{-1}$. The energy available
for dissipation is

\[ E\sim \demi I_{EIZ}(\Omega_{\rm spin}-\Omega_{\rm orb})^2 \]
where $I_{EIZ}$ is the inertia momentum of the zone developing the
elliptic instability. In this first attempt, we shall consider a late type
stars  and restrict, conservatively, the action of this instability to
the convection zone so that it does not interfere with the stratification.
We note that the spin-over modes which
are destabilized are rigid rotations and are thus not affected by the large
turbulent viscosity of the fluid.

Hence, the power dissipated is  approximately $\frac{\eps}{2} I_{\rm
CZ}(\Omega_{\rm spin}-\Omega_{\rm orb})^3$

With these two quantities, one can estimate the time
scale over which synchronization occurs: this is typically the growth of
the elliptical instability. To have orders of magnitude, let us take two
solar type stars orbiting their center of mass in 10 days and let their
spin be twice faster (i.e. $\Omega_{\rm spin}=2\Omega_{\rm orb}$). The
time scale is then 64 years, thus very short.

As far as circularization is concerned, we need to restrict ourselves to
weakly eccentric orbits ($e\la 0.1$ say). Using the
stars to dissipate energy while the angular momentum of the orbit is
assumed constant, the reduction of the eccentricity also occurs on a
short time scale, namely a thousand of years.

The numbers thus derived are rather robust as they depend on geometric
quantities of the system. In view of the observations, which show that
they are systems with a 10 days period which are not circularized nor
synchronous, we may wonder why the elliptic instability is not so
efficient in binary stars. The answer is likely in the much more complex
setup yielded by stars compared to their equivalent in the laboratory.
In stars, there are stratification, magnetic fields, time variable
ellipticity (on eccentric orbits), rotation etc... All these effects
have been explored but most of the time using systems which are not
quite similar to a star. Nevertheless, we shall list them and speculate
about their effects in a stellar situation.

\begin{itemize}
\item {\bf Rotation} of the frame associated with the orbital motion
influences the elliptic instability through the Coriolis force. Craik
(1989) has shown, but using an unbounded strained flow, that rotation is
either stabilizing or destabilizing. It is stabilizing if, for instance,
in an inertial frame, the star does not rotate.

\item {\bf Stratification:} Similarly to rotation, stratification does
not act in a unique sense; using the context of an elliptical cylinder
Kerswell (1993) has shown that stratification is stabilizing for
polar regions of a star but destabilizing for equatorial regions.
\item {\bf Magnetic fields} were found invariably stabilizing (see
Kerswell 1994).
\item {\bf Time-periodic ellipticity} is likely destabilizing through
parametric instabilities. However, the only studied cases are those of
an unbounded strained fluid (Kerswell 2002).
\item {\bf Nonlinear effects} are the most difficult to appreciate.
Experiments have shown that the nonlinear development is violent
(Aldridge et al. 1997) and rapidly leads to a turbulent state (Malkus
1989). In fact, it seems that the saturated state exists only in a very
narrow range of parameters and beyond this range secondary instabilities
due to triadic interactions of inertial modes lead to small scale
motions and a turbulent state (Mason and Kerswell 1999).

Globally, the way the instability saturates may be thought as a change
of the spin axis of the fluid so as to reduce the ellipticity of the
streamlines. In stars such an effect is possible if the spin axis of
stars is inclined in the plane normal to the orbital plane and passing
through the centres of mass. This shows that the apparently inocuous
hypothesis of angular momentum vectors alignment may be crucial to the
instability. It may also be an observational signature. Indeed, we noticed
that the turbulent viscosity of convection zone could hardly inhibit the
resonance of spin-over modes; however, it can easily suppress secondaries
instabilities which, in laboratory experiments lead to a turbulent state.
It may well be that the saturated state is difficult to obtain in the
laboratory but much more easily in  stars.
\end{itemize}

The foregoing discussion shows that many points of the elliptic
instability  in the stellar context remain in the shadows; but because of
its great potential dissipative power, this instability deserves more
study.

\section{Concluding remarks}

Born together, stars of a binary system share the same age and the same initial
metallicity. In most favourable cases, their mass, radii, rotation rates
can be determined. These are of course very interesting pieces for the
puzzle of stellar evolution but there is a price to pay for these
additional informations: the two stars interact during their whole
lifetime. In its simplest form this interaction is of gravitational
origin and generates tides. As we have shown, tides generate various
fluid flows giving rise to transport processes which may be
observationnally constrained if a comparison with analogous isolated
stars is possible.

The foregoing presentation which sketched out all the known mechanisms
by which energy is dissipated also shows that the situation is not simple.
Various processes can dissipate energy and we suggest that among them
the elliptic instability plays a non negligible part. Quite
surprisingly, this instability has been overlooked until now. Despite its
strength, shown by laboratory experiments, observations do not show, yet,
an evidence of this instability. In the list of the mechanisms which may
inhibit this instability, we noticed the misalignement of rotation axis.
As such a misalignement may also result from a saturated state of this
instability and can be an observational signature, it deserves some
study.

\bigskip
\noindent To conclude this contribution let us point out some directions of
research:
\begin{itemize}
\item On the theoretical side more work is obviously needed on the effects
of the elliptic instability.

\item On models, the integration of stellar evolution combined with
dynamical (tidal) evolution following up the works of Claret and Cunha
(1997) and Witte and Savonije (1999b) will be useful to
understand, for instance, the statistical properties of eccentricities as
a function of periods and ages, or the relative importance of pre-main
sequence, main sequence, post-main sequence phases.

\item On the observational side, much  data are needed. First, the
elements describing the dynamics are very much desired: these are the
elements of the orbit $(a,e,i,\Omega,\omega)$, the masses M$_1$, M$_2$,
the spins $\vO_1, \vO_2$ and their variations with time. For instance, the
motion of the apsidal line $\dot{\omega}$ is a quantity constraining the
mass distribution of the stars and therefore their internal rotation.
It is clear that such data will require a lot of efforts but they will help us
much in our understanding of stellar evolution. A pulsar like J0045-7319,
which travels around an early-type star on a highly eccentric orbit, offers
a good step in this direction (Lai et al. 1995).
\end{itemize}

\end{document}